\documentclass[abstract]{scrartcl}

\usepackage[english]{babel}
\usepackage[utf8]{inputenc}
\usepackage[T1]{fontenc}

\usepackage{csquotes}

\usepackage[tt=false]{libertine}
\usepackage{microtype}

\usepackage{xparse}

\usepackage{xcolor}
\definecolor{cite-green}{RGB}{0,115,0}
\definecolor{cau-main}{RGB}{155,10,125}
\usepackage[colorlinks, citecolor=cite-green, linkcolor=cau-main, pdfusetitle]{hyperref}

\usepackage{mathtools, amssymb, amsthm, amsmath, amsfonts, thmtools}
\usepackage{todonotes}

\declaretheorem[numberwithin=section]{theorem}
\declaretheorem[numberwithin=section, sibling=theorem]{lemma}
\declaretheorem[numberwithin=section, sibling=theorem]{corollary}

\usepackage[capitalize,noabbrev]{cleveref}

\newcommand{\vect}[1]{\boldsymbol{#1}}

\newcommand{\ZZ}{\mathbb{Z}}
\newcommand{\QQ}{\mathbb{Q}}
\newcommand{\RR}{\mathbb{R}}

\newcommand{\supp}{\operatorname{supp}}

\newcommand{\bZ}{\mathbb{Z}}

\newcommand{\cT}{\mathcal{T}}

\usepackage{xparse}
\DeclarePairedDelimiter\abs{\lvert}{\rvert}
\DeclarePairedDelimiter\norm{\lVert}{\rVert}
\DeclarePairedDelimiter\floor{\lfloor}{\rfloor}
\DeclarePairedDelimiter\ceil{\lceil}{\rceil}

\DeclarePairedDelimiter\paren{(}{)}
\DeclarePairedDelimiterX\setImpl[1]{\{}{\}}{#1}
\NewDocumentCommand\set{sO{}m>{\TrimSpaces}o}{
  \IfValueTF{#4}{
    \setImpl[#2]{#3 : \IfBooleanTF{#1}{\text{#4}}{#4}}
  }{
    \setImpl[#2]{#3}
  }
}

\newcommand\lex{\ensuremath{\mathrel{<_{\mathrm{lex}}}}}

\usepackage{environ}
\newcommand{\restate}[2]{%
  \begingroup
  \expandafter\renewcommand\csname the#1\endcsname{\ref{#2}}%
  \expandafter\expandafter\expandafter\expandafter\expandafter\expandafter\csname #1\endcsname
  \csname reptheorem@#2\endcsname
  \csname end#1\endcsname
  \endgroup
}

\NewEnviron{restateable}[2]{%
  \global\expandafter\xdef\csname reptheorem@#2\endcsname{%
    \unexpanded\expandafter{\BODY}%
  }%
  \expandafter\expandafter\csname #1\endcsname\BODY\unskip\label{#2}\csname end#1\endcsname
}

\usepackage[
    backend=biber,
    style=alphabetic,
    minalphanames=3,
    maxalphanames=4,
    maxbibnames=99,
    doi=true,
    giveninits=true
]{biblatex}
\title{Canonical Truncation for Mixed-Integer and Combinatorial \(n\)-Fold Optimization}
\author{%
    Klaus Jansen\thanks{Kiel University, kj@informatik.uni-kiel.de, \url{https://orcid.org/0000-0001-8358-6796}}
    \and
    Felix Ohnesorge\thanks{Kiel University, foh@informatik.uni-kiel.de, \url{https://orcid.org/0009-0003-8023-3380}}
    \and
    Björn Schumacher\thanks{Kiel University, bsch@informatik.uni-kiel.de, \url{https://orcid.org/0009-0002-2448-4794}}
}
\date{}
\AfterEndEnvironment{abstract}{%
    \par\smallskip
    {\smallskip\noindent\textbf{Funding.} Deutsche Forschungsgemeinschaft (DFG, German Research Foundation) -- Project number 528381760.}%
}
\begin{document}
\maketitle

\begin{abstract}
  We consider a special block-structured integer program with $n$ blocks.
  Every block has a prescribed number of vectors that need to be selected while minimizing the objective $\vect c^{\intercal}\vect x + g(A\vect x)$, where $c$ is a rational cost vector, $A$ is an integral matrix with $d$ rows, and $g \colon \mathbb{Z}^d \to \mathbb{Q}\cup \{+\infty\}$ is an arbitrary function.
  We access $g$ via an oracle that minimizes $\vect c^{\intercal}\vect x + g(A\vect x)$ when all but $O(d)$ variables are fixed.
  Our algorithm runs in time $(nd\Delta)^{O(d)}(Q + \operatorname{poly}(|I|))$ where $\Delta$ is the largest absolute value of an entry in $A$, $Q$ is the necessary time to evaluate the oracle for $g$ and $|I|$ is the size of the input.
  This model is a variation of the model studied by Dadush et al.~[IPCO 23].
  Their model has only a single block but individual upper and lower bounds on the variables instead of prescribed block sums.
  For their model they present an algorithm with running time $t^d(d\Delta)^{O(d^2)}Q$ where $t$ is the number of variables.

  We use our algorithm to solve mixed-integer linear programs with no upper bounds on the variables, a linear objective, and arbitrarily many constraints involving only fractional variables in time $(m\Delta)^{O(m)}\operatorname{poly}(|I|)$, where $m$ is the number of constraints involving integral variables and $\Delta$ is the largest absolute value of coefficients of integral variables.
  This improves the dependence on $m$ in the exponent to asymptotically match the dependency in the integer setting.
  Further, we use our algorithm to generalize the parameterized algorithm for scheduling on uniformly related machines by Rohwedder~[ICALP 25] to obtain an algorithm for multiple knapsack with running time $w_{\max}^{O(d)}\operatorname{poly}(|I|)$ where $w_{\max}$ is the maximum weight of an item and $d$ the number of different weights.
  This running time is tight in the sense that under the Satisfiability Hypothesis no $w_{\max}^{O(d^{1 - \varepsilon})}\operatorname{poly}(|I|)$ for any $\varepsilon > 0$ is possible.

  % \keywords{Block Structured Integer Programming \and Mixed-Integer Linear Programming \and Non-Linear Optimization}
\end{abstract}

\newpage
\section{Introduction}
We study the following block-structured integer program.
Let \(n,d\geq 1\) be integers.
For each block \(i\), let \(b^{(i)}\) and \(t^{(i)}\) be positive integers, and set \(t=\sum_{i=1}^n t^{(i)}\).
For \(\vect{x}\in\ZZ^t\), let \(\vect{x}^{(i)}\) denote its block of \(t^{(i)}\) consecutive coordinates, starting at \(1+\sum_{h<i}t^{(h)}\).
Let \(A=(A_1\mid\cdots\mid A_n)\in\ZZ^{d\times t}\),
\(\vect{c}\in\QQ^t\), and \(g\colon\ZZ^d\to\QQ \cup \set{+\infty}\).
Set \(\Delta=\max\set{1,\norm{A}_{\infty}}\) where \(\norm{A}_{\infty}\) is the largest absolute value of an entry in \(A\).
We consider
\begin{align}
  \text{minimize}\quad  & \vect{c}^{\intercal}\vect{x} + g(A\vect{x}) \notag                                   \\
  \text{such that}\quad & \mathbf{1}^{\intercal}\vect{x}^{(i)}=b^{(i)} \quad (i\in[n]),\label{eq:main-problem} \\
                        & \vect{x}\in\ZZ_{\geq 0}^t.\notag
\end{align}
We optimize for the combination of \(\vect c^{\intercal}\vect x\) and \(g\) applied to the linking image \(A\vect x\) while every prescribed block sum, i.e., \(\mathbf{1}^{\intercal}\vect x^{(i)} = b^{(i)}\), is satisfied.
We are given \(g\) via an oracle with the following low-dimensional optimization assumption.
Given a partition \(K \sqcup J\) of \([t]\), the coordinates with indices in \(J\) are fixed and determined by the vector \(\vect z\in\bZ_{\geq 0}^J\).
Further, we have \(\abs{K} \leq 6d - 3\).
The oracle solves
\begin{align}
  \text{minimize}\quad  & \vect{c}_K^{\intercal} \vect{x} + \vect c_J^{\intercal}\vect{z} + g(A_K\vect{x} + A_J\vect{z}) \notag \\
  \text{such that}\quad & \mathbf{1}^{\intercal}\vect{x}^{(i)} + \mathbf{1}^{\intercal}\vect{z}^{(i)} =b^{(i)}
  \quad (i\in[n]),\label{eq:oracle}                                                                                             \\
                        & \vect{x}\in\ZZ_{\geq 0}^K,\notag
\end{align}
where \(c_K, c_J\) and \(A_K, A_J\) are the restrictions to the columns given by \(K\) and \(J\) respectively.
\(\vect x^{(i)}\) and \(\vect z^{(i)}\) contain the coordinates of \(K\) and \(J\), respectively, belonging to the original block \(i\).
The oracle reports infeasibility in case no feasible integer point exists due to the supplied \(J\) and \(\vect z\).
If the optimum is \(+\infty\), it is returned symbolically.
Let \(Q\geq 1\) be the necessary time to evaluate the oracle \labelcref{eq:oracle}, return the optimal value, and a corresponding solution.

\begin{restateable}{theorem}{thm:main}
  Under the oracle assumption in \labelcref{eq:oracle},
  the problem \labelcref{eq:main-problem} can be solved in
  \((nd\Delta)^{O(d)}(Q + \operatorname{poly}(\abs{I}))\) time.
\end{restateable}
An evaluation oracle for an arbitrary \(g\) does not suffice for the claimed bound.
When \(n=d=1\), \(A=(0,1)\), \(\vect{c}=\vect{0}\), and \(b^{(1)}=b\), the problem is \(\min_{z\in\{0,\ldots,b\}}g(z)\), which requires \(\Omega(b)\) evaluation oracle calls for arbitrary functions.

Our oracle model is an adaptation of the one by Dadush et al.~\cite{DBLP:conf/ipco/DadushLRV23}.
They optimize \(\vect{c}^{\intercal}\vect{x}+g(W\vect{x})\) with lower and upper bounds for \(\vect x\), where \(W\in\ZZ^{d\times t}\) has entries bounded by \(\Delta\), assuming that optimization is possible when all but \(d\) variables are fixed.
Their algorithm has running time \(t^d(d\Delta)^{O(d^2)}Q\), where \(Q\) again is the cost of the oracle.
The factor \(t^d\) can be absorbed into the parameter dependence in several special cases.
% We introduce multiple blocks and replace the upper and lower bounds by prescribed block sums and obtain \((nd\Delta)^{O(d)}\) dependence under the oracle assumption \ref{eq:oracle}.
Hunkenschröder et al.~\cite{DBLP:journals/siamjo/HunkenschroderP23} considered the 0/1 variant with the restriction of \(g\) to special convex functions.
Their assumed oracle provides the function value and the gradient at the queried point.
They obtain an \((m\Delta)^{O(m^3)}\operatorname{poly}(\abs{I})\) algorithm even when \(W\) is unknown under certain conditions.

\subsection{Applications}
\subsubsection{Mixed-Integer Linear Programming}
Let \(A\in\ZZ^{m\times p}\), \(C\in\QQ^{m\times q}\), \(\vect{b}\in\QQ^m\), \(\vect{c}\in\QQ^p\), and \(\vect{e}\in\QQ^q\), where \(m,p\geq 1\).
Let \(P\subseteq\RR^q\) be an explicitly given rational polyhedron, and set \(\Delta=\max\set{1,\norm{A}_{\infty}}\).
We consider the following mixed-integer linear program
\begin{align}
  \text{minimize}\quad  & \vect{c}^{\intercal}\vect{x}+\vect{e}^{\intercal}\vect{y} \notag \\
  \text{such that}\quad & A\vect{x}+C\vect{y}=\vect{b} \label{eq:milp}                     \\
                        & \vect{x}\in\ZZ_{\geq 0}^p,\quad \vect{y}\in P\notag
\end{align}
To solve \labelcref{eq:milp}, we apply \Cref{thm:main} for a single block together with a calculated block sum while encoding the fractional part of \labelcref{eq:milp} in \(g\) (as done in \cite{DBLP:conf/ipco/DadushLRV23}).
We solve the corresponding oracle using the algorithm of Lenstra with Kannan's improvement~\cite{Lenstra1983,Kannan1987}, which solve MILPs in time \(p^{O(p)}\operatorname{poly}(\abs{I})\).
This  yields the following result.
\begin{restateable}{theorem}{thm:milp}
  The MILP \labelcref{eq:milp} can be solved in  \((m\Delta)^{O(m)}\operatorname{poly}(|I|)\) time.
\end{restateable}
Dadush et al.~\cite[Corollary~3]{DBLP:conf/ipco/DadushLRV23} already consider a closely related mixed-integer model with upper bounds on the integer variables, arbitrary coefficients on the continuous variables, and additional constraints involving only those variables.
Their bound is \(p^m(m\Delta)^{O(m^2)}\operatorname{poly}(|I|)\).
Without integer-variable upper bounds or the linear objective, the \(p^m\) factor can be omitted.
The running time \((m\Delta)^{O(m^2)}\operatorname{poly}(|I|)\) for the MILP \labelcref{eq:milp} was also achieved in \cite{DBLP:phd/dnb/Rohwedder19}.
Thus, our contribution is reducing the dependence on \(m\) in the exponent, which asymptotically matches the best known dependence on \(m\) in the exponent for pure non-negative integer linear programming \cite{EisenbrandWeismantel2020,JansenRohwedder2023}.
Brand et al.~\cite{BrandEtAl2024} give algorithms for separable convex mixed-integer optimization with few rows, with parameter dependence \((m\norm{E}_\infty)^{O(m^2)}\) for the full constraint matrix \(E\), where \(m\) is the number of rows in \(E\).
Compared to \labelcref{eq:milp}, extra constraints for the fractional variables outside of \(E\) are not permitted in their model.

\subsubsection{Combinatorial \(n\)-fold}
Let \(A=(A_1\mid\cdots\mid A_n)\), \(\vect{c}\), and the positive block totals \(b^{(i)}\) be as in \labelcref{eq:main-problem}, and let \(\vect{u}\in\QQ^d\).
Let \(f \colon \bZ^d \to \QQ\) be a separable discrete convex function, i.e., \(f_h(s+1)-f_h(s)\) is non-decreasing in \(s\) for every \(h\in[d]\).
% For each \(h=1,\ldots,d\), let \(f_h\colon\ZZ\to\QQ\) be discrete convex, meaning that \(f_h(s+1)-f_h(s)\) is nondecreasing in \(s\).
We consider the combinatorial \(n\)-fold problem
\begin{equation}\label{eq:combinatorial-nfold}
  \begin{aligned}
    \text{minimize}\quad & \vect{c}^{\intercal}\vect{x}+ f(A\vect x)  \\
    \text{such that}\quad
                         & A\vect{x}\leq \vect{u},                    \\
                         & \vect{1}^{\intercal}\vect{x}^{(i)}=b^{(i)}
    \quad(i=1,\ldots,n),                                              \\
                         & \vect{x}\in\ZZ_{\geq 0}^t.
  \end{aligned}
\end{equation}
Knop et al.~\cite{KnopKouteckyMnich2020} introduced the combinatorial \(n\)-fold model that specializes \(n\)-fold IPs by restricting the block-local matrices to rows containing only ones.
Our addition is the separable convex cost on the linking image \(A\vect{x}\).
\begin{restateable}{theorem}{thm:combinatorial-nfold}
  The combinatorial \(n\)-fold \labelcref{eq:combinatorial-nfold} can be solved in \((nd\Delta)^{O(d)}\operatorname{poly}(|I|)\) time, assuming that \(f\) can be evaluated exactly in time \(\operatorname{poly}(|I|)\).
\end{restateable}
For linear objectives and a fixed linking right-hand side, Jansen et al.~\cite{JansenEtAl2025} already obtain \((nd\Delta)^{O(d)}\) parameter dependence with logarithmic dependence on the right-hand side.
The same asymptotic dependency on \(n\), \(d\), and \(\Delta\) was also achieved by Rohwedder~\cite{Rohwedder2025} and Gupta et al.~\cite{GuptaEtAl2025} but both algorithms have a linear dependence on the total prescribed block sums.
None of the just mentioned algorithms are able to handle non-linear objectives.
Notable about this line of research is that the quadratic dependence on the number of linking constraints in the exponent of \(n\)-fold algorithms (e.g.\ \cite{CslovjecsekEtAl2021}) is replaced by introducing the number of blocks into the base of the parameter term.

\subsubsection{Multiple Knapsack}
An instance of Multiple Knapsack consists of item types with positive integer weights \(w_i\), nonnegative rational profits \(p_i\),
and integer multiplicities \(n_i\geq 0\), together with \(m\) knapsacks of integer capacities \(C_1,\ldots,C_m > 0\).
The task is to select and assign items to the knapsacks without exceeding their capacities or the available multiplicities, maximizing the total profit.
All numbers, including the multiplicities, are encoded in binary.
Let \(d\) be the number of distinct item weights and let \(w_{\max}=\max_i w_i\) be the largest weight.
A running time of \(w_{\max}^{O(d^2)}\operatorname{poly}(|I|)\) already follows directly from a straightforward \(n\)-fold IP formulation with classic convex costs and standard \(n\)-fold algorithms~\cite{DBLP:journals/mp/KnopKLMO23}: the formulation has \(d\) linking constraints for the weight classes, one local capacity constraint per knapsack, and an additional block recording rejected items and their profits.

We remove the quadratic dependence on \(d\) in the exponent by combining our combinatorial \(n\)-fold algorithm with a modular reduction for scheduling on uniformly related machines by Rohwedder~\cite{Rohwedder2025}.
\begin{restateable}{theorem}{thm:multiple-knapsack}
  Multiple Knapsack can be solved in time \(w_{\max}^{O(d)}\cdot \operatorname{poly}(|I|)\).%, where \(|I|\) is the binary encoding length of the instance.
\end{restateable}
Note that the dependence on \(w_{\max}\) cannot be removed: the reduction of Kouteck\'y and Zink~\cite[Theorem~4.13]{KouteckyZink2025} for scheduling on uniformly related machines implies that Multiple Knapsack is NP-hard already for \(d=6\) and unit profits, and hence is not in XP parameterized only by \(d\) unless \(\mathrm{P}=\mathrm{NP}\).
% Indeed, scheduling feasibility with makespan threshold \(T\) and machine speeds \(s_j\) reduces to packing all items into knapsacks of capacities \(\floor{Ts_j}\), preserving the distinct weights and their multiplicities.
% Moreover, combining their reduction~\cite[Lemmas~4.8 and~4.11]{KouteckyZink2025} with the bin-packing lower bound of Bringmann, D\"urr, and W\k{e}grzycki~\cite{BringmannDuerrWegrzycki2026} rules out \(f(d)w_{\max}^{o(d)}\operatorname{poly}(|I|)\) time under the Exponential Time Hypothesis (ETH), for every computable function \(f\), even with unit profits. Thus, the linear dependence on \(d\) in the exponent is optimal up to constant factors under ETH.
A lower bound follows under the Satisfiability Hypothesis (SH) of Stearns and Hunt~\cite{StearnsHunt1990}:
for every fixed \(\varepsilon>0\), there is no \(w_{\max}^{O(d^{1-\varepsilon})}\cdot \operatorname{poly}(|I|)\)-time algorithm, even with identical capacities and unit profits.
Indeed, the reduction of Chen, Jansen, and Zhang~\cite{ChenJansenZhang2014} maps a 3SAT instance of size \(s\) to \(O(s)\) jobs with polynomially bounded processing times.
Such an algorithm would therefore solve 3SAT in time \(2^{O(s^{1-\varepsilon}\log s)}\leq 2^{O(s^{1-\varepsilon/2})}\), contradicting SH.

Other parameterizations and approximation algorithms have also been studied for the classical formulation with explicitly listed items (see e.g., \cite{ChekuriKhanna2005,GurskiRehsRethmann2019,Jansen2009,Jansen2012}).
% Gurski, Rehs, and Rethmann~\cite{GurskiRehsRethmann2019} investigate parameterized algorithms and kernelization, including parameterizations by the number of items and by the combined number of knapsacks and maximum capacity.
% For arbitrary knapsack capacities, Chekuri and Khanna~\cite{ChekuriKhanna2005} give a polynomial-time approximation scheme (PTAS), and Jansen~\cite{Jansen2009} gives an efficient polynomial-time approximation scheme (EPTAS), subsequently improving its running time~\cite{Jansen2012}.

\subsection{Technical Overview}
The core idea of our algorithm is the classic paradigm to focus on optimal solutions with special structure.
Since the evaluation of \(g\) only depends on the linking image, the concrete assignment of the variables only matters for the linear objective \(\vect c\).
For a fixed linking image, we choose a representative candidate from the set of optimal solutions according to \(\vect c\).
In particular, we choose the lexicographically (lexicographic order denoted by \(\lex\)) smallest solution.
We call these solutions canonical.
Considering the lexicographically smallest solution is a common approach, see for example \cite{DBLP:conf/ipco/RohwedderW25}.
Therefore, our contribution is applying this technique in our proposed setting and its application.

Canonical solutions have inherently small support: the number of non-zero coordinates in every block is bounded by a function in \(d\) and \(\Delta\).
Further, the number of blocks with multiple big coordinates (coordinates exceeding a certain threshold) and the total number of big coordinates in these blocks is bounded by \(O(d)\).
The proofs for the structural properties rely on the fact that the variables have no individual upper bounds.
These structural properties allow us to define truncated canonical solutions, which can be efficiently enumerated while retaining all the necessary information to invoke the oracle.
This is in contrast to the approach used in \cite{DBLP:conf/ipco/DadushLRV23}, which builds on the proximity for ILPs in \cite{EisenbrandWeismantel2020} and combines it with a guessing procedure to apply the oracle.

\section{Properties of Canonical Solutions}
We define
\begin{equation*}
  p \colon \ZZ_{\geq 0}^t \to \ZZ_{\geq 0}^t, \vect{u} \mapsto \operatorname{lexmin}\operatorname*{arg\,min}_{\vect{v}\in\ZZ_{\geq 0}^t}
  \set[\Bigg]{\vect{c}^{\intercal}\vect{v}}[
    \begin{aligned}
       & A\vect{v}=A\vect{u},                                                                 \\
       & \mathbf{1}^{\intercal}\vect{u}^{(i)}=\mathbf{1}^{\intercal}\vect{v}^{(i)}\ (i\in[n])
    \end{aligned}].
\end{equation*}
The function is well-defined, since \(\vect u\) is contained in the finite candidate set for \(p(\vect u)\).
Call \(\vect{x}\) \emph{canonical} if it is a fixpoint of \(p\), i.e., \(p(\vect{x}) = \vect{x}\).
Replacing an optimal solution \(\vect{x}^\star\) by its canonical representative \(p(\vect{x}^\star)\) preserves all constraints and cannot increase its linear cost.
Consequently a canonical optimum exists.

First, we show that being canonical is hereditary in the sense that all subvectors of a canonical vector are also canonical.
\(\vect u\in\bZ^t\) is a subvector of \(\vect x\in\bZ^t\) if \(u_i \leq x_i\) for all \(i\in[t]\).
We denote this by \(\vect u \leq \vect x\).
\begin{lemma}\label{prop:canonicity-is-hereditary}
  Let \(\vect{x}\in\ZZ_{\geq 0}^t\) be canonical.
  Then every \(\vect{u}\in\ZZ_{\geq 0}^t\) with \(\vect{u} \leq \vect{x}\) is also canonical.
\end{lemma}
\begin{proof}
  Let \(\vect u\in\bZ_{\geq 0}^t\) such that \(\vect u \leq \vect x\).
  Towards a proof by contradiction, suppose \(\vect{v} \coloneq p(\vect{u}) \ne \vect{u}\).
  Set \(\vect{x}' = \vect{x} + \vect{v} - \vect{u}\).
  Since \(A\vect{v} = A\vect{u}\) and \(\mathbf{1}^{\intercal}\vect{v}^{(i)} = \mathbf{1}^{\intercal}\vect{u}^{(i)}\) for all \(i\in[n]\), we have \(A\vect{x}' = A\vect{x}\) and \(\mathbf{1}^{\intercal}\vect{x}'^{(i)} = \mathbf{1}^{\intercal}\vect{x}^{(i)}\) for all \(i\in[n]\).
  If \(\vect{c}^{\intercal}\vect{v} < \vect{c}^{\intercal}\vect{u}\) we have \(\vect{c}^{\intercal}\vect{x}' < \vect{c}^{\intercal}\vect{x}\).
  Otherwise, if \(\vect{c}^{\intercal}\vect{v} = \vect{c}^{\intercal}\vect{u}\) and \(\vect{v} \lex \vect{u} \) we have \(\vect{c}^{\intercal}\vect{x}' = \vect{c}^{\intercal}\vect{x}\) and \(\vect{x}' \lex \vect{x}\).
  Both cases contradict that \(\vect{x}\) is a fixpoint of \(p\), which implies \(\vect{v} = \vect{u}\).
\end{proof}
A direct implication of this lemma is that all subvectors of a canonical solution have a different linking sum or block sum.
\begin{corollary}\label{prop:canonic-image-is-different}
  Let \(\vect{x}\in\ZZ_{\geq 0}^t\) be canonical.
  Then, for \(\vect u, \vect v\in\bZ_{\geq 0}^t\) with \(\vect u, \vect v \leq \vect x\), \(A\vect u = A\vect v\) and \(\mathbf{1}^{\intercal}\vect u^{(i)} = \mathbf{1}^{\intercal}\vect v^{(i)}\) for all \(i\in[n]\) implies \(\vect u = \vect v\).
\end{corollary}
\begin{proof}
  Let \(\vect u, \vect v\in\bZ_{\geq 0}^t\) with \(\vect u, \vect v \leq \vect x\), \(A\vect u = A\vect v\) and \(\mathbf{1}^{\intercal}\vect u^{(i)} = \mathbf{1}^{\intercal}\vect v^{(i)}\) for all \(i\in[n]\).
  Thus, the candidate set for \(p(\vect v)\) and \(p(\vect u)\) is the same and since \(p(\vect u) = \vect u\) and \(p(\vect v) = \vect v\), we have \(\vect u = \vect v\).
\end{proof}
This fact is important for the structural properties we show next since it enables counting arguments.
We start by showing that a canonical solution inherently has small support in every block.
The support of a vector \(\vect x\in\RR^t\) is defined as \(\supp(\vect x) \coloneqq \set{i\in[t]}[x_i \ne 0]\).
\begin{lemma}\label{prop:support}
  Let \(\vect{x}\in\ZZ_{\geq 0}^t\) be canonical.
  Then
  \[
    \abs{\supp(\vect{x}^{(i)})} < S \coloneqq 8(d+1)^2\Delta \qquad \text{for all } i\in[n].
  \]
\end{lemma}
\begin{proof}
  Towards a proof by contradiction, suppose that there is an \(i\in[n]\) such that \(\abs{\supp(\vect{x}^{(i)})} \geq S \coloneqq 8(d+1)^2\Delta\).
  Set \(B \coloneqq \set{\vect{u}\in\set{0, 1}^t}[\vect{u} \leq \vect{x}, \norm{\vect{u}}_1 \leq S, \vect{u}^{(h)}=\vect{0} \text{ for } h\ne i]\) with \(\abs{B} \geq 2^S\).
  For every \(\vect{u}\in B\) we have \(\norm{A\vect{u}}_{\infty} \leq S\Delta\) and \(\norm{\vect{u}}_1 \leq S\).
  Hence, the number of possible images \((A\vect{u},\norm{\vect{u}}_1)\) is bounded by \((2S\Delta + 1)^{d + 1}\).
  Since \(2^m>m+1\) for every integer \(m\geq 2\), we have
  \[
    \abs{B} \geq 2^S
    = \paren{2^{4(d+1)\Delta}}^{2(d+1)}
    > \paren{4(d+1)\Delta+1}^{2(d+1)}
    \geq (2S\Delta+1)^{d+1},
  \]
  where the last inequality uses \((4(d+1)\Delta+1)^2\geq 16(d+1)^2\Delta^2+1=2S\Delta+1\).
  Thus two different vectors in \(B\) have the same linking image and all block sums, which contradicts \cref{prop:canonic-image-is-different}.
\end{proof}
We call the coordinate \(r\) in a solution \(\vect{x}\) \emph{big} if \(x_r \geq q \coloneqq 4d\Delta\) and \emph{small} otherwise.
The set of big coordinates for block \(i\) is denoted by
\[
  B^{(i)}(\vect{x})=\{j:x_j^{(i)}\geq q\}.
\]
If a block has only a single big coordinate its value is fully determined by the assignment of the small variables in the block and \(b^{(i)}\).
Under the assumption that we know the values of all small coordinates, these can be determined in the algorithm without invoking the oracle.
Thus, we are only interested in the total number of big coordinates in blocks with at least two big coordinates.
This number is in \(O(d)\) as shown by the following lemma.
\begin{lemma}\label{prop:big-vars}
  Let \(\vect{x}\in\ZZ_{\geq 0}^t\) be canonical.
  Then
  \[
    \sum_{\substack{i\in[n]\\\abs{B^{(i)}(\vect x)} \geq 2}} \abs{B^{(i)}(\vect x)} \leq 6d - 3.
  \]
\end{lemma}
\begin{proof}
  Suppose there are \(2d\) disjoint pairs \((p_i, h_i)\) of big coordinates in \(\vect x\), with each pair contained in a single block.
  We build \((q + 1)^{2d}\) different subvectors of \(\vect x\).
  Given a vector \(\vect y\in\set{0, \ldots, q}^{2d}\) set \(u_{p_i} \coloneqq y_i\) and \(u_{h_i} = q - y_i\) for all \(i\in[2d]\).
  Set all other coordinates to zero and call this vector \(\vect u_{\vect y}\).
  All these vectors are subvectors of \(\vect x\) and hence canonical.
  Further, all constructed vectors share the same block sums and \(\norm{A\vect u_{\vect y}}_{\infty} \leq 2dq\Delta\).
  Thus there are at most \((4dq\Delta + 1)^d\) different linking images but \((4dq\Delta+1)^d=(q^2+1)^d<(q+1)^{2d}\) which contradicts \cref{prop:canonic-image-is-different}.

  The maximum number of distinct disjoint pairs in block \(i\) is \(\floor{\frac{\abs{B^{(i)}(x)}}{2}}\).
  It follows that
  \[
    \sum_{\substack{i\in[n]\\\abs{B^{(i)}(\vect x)} \geq 2}} \floor[\Big]{\frac{\abs{B^{(i)}(\vect x)}}{2}} \leq 2d - 1.
  \]
  Since \(k \leq 3\floor{\frac{k}{2}}\) for \(k \geq 2\), we obtain the claimed bound.
\end{proof}
Next, we truncate all big coordinates in canonical vectors such that we still retain the information which coordinates are big and are able to enumerate all truncated vectors efficiently.
\begin{lemma}\label{prop:truncated-solution}
  Let \(\vect x\in\bZ_{\geq 0}^t\) be canonical with \(\norm{\vect x^{(i)}}_1 = b^{(i)}\) for \(i\in[n]\).
  Then there exists a truncation \(\vect u\in\bZ_{\geq 0}^t\) of \(\vect x\) such that
  \begin{enumerate}
    \item \(\vect u \leq \vect x\), hence, by heredity \(\vect u\) is canonical,
    \item \(\mathbf{1}^{\intercal}\vect u^{(i)} = \overline{b}^{(i)} \coloneqq \min\set{b^{(i)}, qS}\),
    \item \(\norm{A\vect u}_{\infty} \leq nqS\Delta\), and
    \item \(x_j > u_j \implies u_j \geq q\) for all \(j\in[t]\).
  \end{enumerate}
\end{lemma}
\begin{proof}
  We construct \(\vect u\) from \(\vect x\).
  Start with \(\vect u\leq \vect x\) where the entries are given by \(u_j=\min\{x_j,q\}\).
  Its sum in block \(i\) is less than \(qS\) by \cref{prop:canonicity-is-hereditary,prop:support}, and at most \(b^{(i)}\).
  Increase its coordinates integrally, never exceeding \(\vect x\), until \(\mathbf{1}^{\intercal}\vect u^{(i)} = \overline b^{(i)}\) in every block.
  This is possible since \(\overline b^{(i)} \leq b^{(i)}\).
  The resulting vector satisfies all claimed properties.
\end{proof}

\section{Algorithm}
The algorithm works by first enumerating all truncated candidates as defined by \cref{prop:truncated-solution}.
Then for every candidate we fix all fully determined variables and solve the remaining problem with the oracle given in \labelcref{eq:oracle}.

We start by eliminating duplicate columns in every block.
Within each block, merge variables whose columns in \(A_i\) agree, retaining one of minimum linear cost.
Since we only have prescribed block sums and no individual lower/upper bounds for variables, this transformation preserves all the necessary structure.
After this step, the number of variables is bounded by \(n(2\Delta+1)^d\).
In the following we use the same notation for the reduced instance.

Define the set of truncated candidates with the bounds given in \Cref{prop:truncated-solution}
\[
  \cT \coloneqq \set[\big]{\vect u\in\bZ_{\geq 0}^t}[p(\vect u) = \vect u, \norm{A\vect u}_{\infty} \leq nqS\Delta, \mathbf{1}^{\intercal}\vect u^{(i)} =\overline b^{(i)} (i\in[n])].
\]
This set can be enumerated efficiently using straightforward dynamic programming.
\begin{lemma}\label{prop:enumerate-truncated-candidates}
  \(\cT\) can be enumerated in \((2nd\Delta)^{O(d)}\operatorname{poly}(\abs{I})\) time.
\end{lemma}
\begin{proof}
  We use a dynamic program processing blocks in order and inserting \(\overline b^{(i)}\) units into block \(i\), one at a time.
  At layer \((i,h)\), for \(0\leq h\leq\overline b^{(i)}\), let \(P_{i,h}(\vect \rho)\) store the minimum-cost, lexicographically smallest vector with linking image \(\vect \rho\), block sums \(\overline b^{(\ell)}\) for \(\ell<i\), sum \(h\) in block \(i\), and zero in later blocks.
  Start with the zero vector at image zero.
  The initial layer of each block is the preceding block's final layer, and for \(h \geq 1\) the recurrence is
  \begin{equation*}
    P_{i,h}(\vect \rho)=
    \operatorname{lexmin}\operatorname*{arg\,min}_{\vect u\in\bZ_{\geq 0}^t}
    \set[\big]{\vect c^{\intercal}\vect u}[\vect u = P_{i,h-1}(\vect \rho-A_{ij})+\vect e_{ij}, 1\leq j\leq t^{(i)}],
  \end{equation*}
  where \(A_{ij}\) is the \(j\)th column of \(A\) in block \(i\) and \(\vect e_{ij}\) is the unit vector corresponding to the \(j\)th variable in block \(i\).
  Removing a positive coordinate from any candidate gives a predecessor.
  Replacing this predecessor by the stored minimum-cost, lexicographically smallest representative cannot worsen the resulting vector.
  Induction therefore proves the recurrence.

  The preprocessing can be done in time \(\operatorname{poly}(\abs{I})\).
  For all states \(\vect \rho\), we have \(\norm{\vect \rho}_{\infty}\leq nqS\Delta\) since \(\norm{P_{i, h}(\vect \rho)}_1 \leq nqS\).
  Every state has at most \((2\Delta + 1)^d\) potential predecessors and thus a state can be calculated in \(O((d + t) \cdot (2\Delta + 1)^d)\).
  Since \(\cT\) is the last layer of the dynamic program, it can be enumerated in time
  \begin{equation*}
    nqS \cdot (2nqS\Delta + 1)^d \cdot O((d + t) \cdot (2\Delta + 1)^d)\cdot\operatorname{poly}(\abs{I}) \leq (2nd\Delta)^{O(d)} \operatorname{poly}(\abs{I}). \tag*{\qedhere}
  \end{equation*}
\end{proof}

\restate{theorem}{thm:main}
\begin{proof}
  If \(nd\Delta=1\), then \(n=d=\Delta=1\), and every column of \(A\) belongs to \(\{-1,0,1\}\).
  Retain a minimum-cost variable for each column value and fix all other variables to zero.
  This preserves an optimum and leaves at most \(3=6d-3\) free variables, so a single oracle call solves the problem in \(Q + \operatorname{poly}(\abs{I})\) time.
  Henceforth assume \(nd\Delta\geq 2\).

  The remaining part of the algorithm is to apply the oracle.
  For each truncated candidate \(\vect u\in\cT\), we define a set \(J\subseteq [t]\) and vector \(\vect z\in\bZ_{\geq 0}^J\) to supply to the oracle.
  \(J\) is the set of all indices corresponding to small coordinates in \(\vect u\), i.e., \(u_i < q\), and all indices of big coordinates that are the only big coordinate in their block.
  For all indices \(i\) corresponding to small coordinates, we set \(z_i = u_i\).
  Let \(j\) be the index of the single big coordinate in block \(i\).
  We set \(z_j = b^{(i)} - (\norm{\vect u^{(i)}}_1 - u_j)\).
  Finally set \(K \coloneqq [t]\setminus J\).
  We have \(\abs{K} \leq 6d - 3\) by \Cref{prop:big-vars}.
  We call the oracle \labelcref{eq:oracle} with \((K, J)\) and \(\vect z \in\bZ_{\geq 0}^J\).
  Discard the result if the oracle reports infeasibility.
  Return the best solution over all candidates, which may be \(+\infty\) when all feasible oracle calls return this value.

  Let \(\vect x^{\star}\) be an optimal canonical solution.
  By \cref{prop:truncated-solution} and the definition of \(\cT\) the truncated version \(\overline{\vect x}\) of \(\vect x^{\star}\) will be considered by the algorithm.
  Let \(J\subseteq[t]\) and \(\vect z\) be the set and vector constructed in the iteration of \(\overline{\vect x}\).
  We have \(z_i = x^{\star}_i\) for all \(i\in J\) by property 4 of \cref{prop:truncated-solution} and the fact that the value for a unique big coordinate in a block is fully determined by the values of the small variables.
  Thus, the oracle in this iteration finds a solution at least as good as \(\vect x^{\star}\).

  The preprocessing and enumerating the set \(\cT\) takes time \((2nd\Delta)^{O(d)}\operatorname{poly}(\abs{I})\) by \cref{prop:enumerate-truncated-candidates}.
  The number of oracle calls is bounded by \((2nd\Delta)^{O(d)}\).
  Since \(2nd\Delta\leq(nd\Delta)^2\) when \(nd\Delta\geq 2\), we obtain the claimed running time. 
\end{proof}

\section{Applications}
\subsection{MILP}

\restate{theorem}{thm:milp}

\begin{proof}
  We start by solving the LP relaxation of \labelcref{eq:milp}.
  If the LP relaxation is infeasible, then so is the MILP, and we are done.
  Since all data are rational, if the LP relaxation has an optimal solution, the MILP is either infeasible or has an optimal solution; if the LP relaxation is unbounded below, the MILP is either infeasible or also unbounded below~\cite{DBLP:journals/mp/Meyer74}.

  Suppose first that the LP relaxation has an optimal solution.
  For fixed \(\vect{x}\), the constraints on \(\vect{y}\) depend
  on \(\vect{x}\) only through \(\vect{\rho}=A\vect{x}\). To express the MILP in the form
  required by \cref{thm:main}, define, for \(\vect{\rho}\in\ZZ^m\),
  \begin{equation*}
    g(\vect{\rho})=
    \begin{cases}
      \min\{\vect{e}^{\intercal}\vect{y}:C\vect{y}=\vect{b}-\vect{\rho},\ \vect{y}\in P\},
               & \text{if the feasible set is nonempty}, \\
      +\infty, & \text{otherwise}.
    \end{cases}
  \end{equation*}
  All nonempty LPs defining \(g\) have the same recession directions, namely those of \(P\) that lie in \(\ker C\).
  An improving direction would also make the original LP relaxation unbounded below, so every nonempty LP defining \(g\) has a finite, attained rational minimum.
  The MILP therefore reduces to minimizing \(\vect{c}^{\intercal}\vect{x}+g(A\vect{x})\) over \(\vect{x}\in\ZZ_{\geq 0}^p\), where \(g(A\vect{x})=+\infty\) excludes choices of \(\vect{x}\) that admit no feasible \(\vect{y}\).

  To apply \cref{thm:main} with one block and \(d=m\), we also need a prescribed sum of the integer coordinates.
  We will obtain this by bounding \(\vect{1}^{\intercal}\vect{x}\) and adding a slack variable.
  The bound must preserve an optimal solution whenever one exists and have polynomial binary encoding length.
  Let \((\overline{\vect x},\overline{\vect y})\) be the optimal LP solution computed above, chosen with polynomial encoding length.
  Let \(B\) be the integer constraint matrix obtained by writing the full constraint system as inequalities and clearing denominators row by row, and set \(N=p+q\) and \(H=\max\set{1,\norm{B}_{\infty}}\).
  Its square subdeterminants have absolute value at most \(N!H^N\).
  Thus, by the mixed-integer proximity bound in \cite[Remark~1]{DBLP:journals/mp/CookGST86}, converting from the infinity norm to the \(1\)-norm, whenever the MILP is feasible it has an optimal solution \((\vect x^*,\vect y^*)\) with
  \(\norm{(\overline{\vect x},\overline{\vect y})-(\vect x^*,\vect y^*)}_1 \leq U\coloneqq N^2N!H^N\).

  The positive integer \(U\) is computable in polynomial time and has polynomial encoding length.
  Set \(b^{(1)}\coloneqq\ceil{\norm{\overline{\vect x}}_1}+U\).
  Then \(b^{(1)}\) has polynomial encoding length and \(\mathbf{1}^{\intercal}\vect x^*\leq b^{(1)}\).
  Denominator clearing is used only to derive this bound; the reduction retains the original linking matrix \(A\).

  Introduce \(s\in\ZZ_{\geq 0}\) with zero cost and a zero linking column, so that \(\vect{1}^{\intercal}\vect{x}+s=b^{(1)}\).
  Write \(\widetilde A=(A\mid\vect 0)\), \(\widetilde{\vect c}=(\vect c,0)\), and \(\vect w=(\vect x,s)\).
  We obtain
  \begin{equation*}
    \min\set[\Big]{\widetilde{\vect c}^{\intercal}\vect w+g(\widetilde A\vect w)}[
      \mathbf{1}^{\intercal}\vect w=b^{(1)},\quad
      \vect w\in\ZZ_{\geq 0}^{p+1}].
  \end{equation*}
  This has one block, \(m\) linking rows, and coefficient bound \(\Delta\).
  It has a finite optimum exactly when the MILP \labelcref{eq:milp} is feasible, and then preserves its optimal value.

  It remains to implement the oracle in \labelcref{eq:oracle} for this choice of \(g\).
  For a given partition \(K\sqcup J=[p+1]\) with \(\abs{K}\leq 6m-3\) and \(\vect z\in\ZZ_{\geq 0}^J\), first check whether the block-sum constraint admits a nonnegative integer completion, reporting infeasibility otherwise.
  Substituting the definition of \(g\) then gives the MILP
  \begin{equation*}
    \begin{aligned}
      \text{minimize}\quad  & \widetilde{\vect c}_K^{\intercal}\vect w+\vect{e}^{\intercal}\vect y \\
      \text{such that}\quad & \widetilde A_K\vect w+\widetilde A_J\vect z+C\vect y=\vect b,        \\
                            & \mathbf{1}^{\intercal}\vect w+\mathbf{1}^{\intercal}\vect z=b^{(1)}, \\
                            & \vect w\in\ZZ_{\geq 0}^K,\quad\vect y\in P.
    \end{aligned}
  \end{equation*}
  If this MILP is infeasible, the oracle returns \(+\infty\) together with any nonnegative integer completion of the block sum.
  Otherwise, it returns an optimal integer vector and the corresponding objective value.
  In all calls generated by the algorithm, the fixed coordinates are at most \(b^{(1)}\), so the oracle inputs have polynomial encoding length.
  Since this MILP has only \(O(m)\) integer variables, it can be solved in time \(Q=m^{O(m)}\operatorname{poly}(|I|)\)~\cite{Lenstra1983,Kannan1987}; the continuous dimension affects only the polynomial factor.
  Applying \cref{thm:main} now gives
  \[
    (m\Delta)^{O(m)}\bigl(Q+\operatorname{poly}(|I|)\bigr)
    =(m\Delta)^{O(m)}\operatorname{poly}(|I|).
  \]
  If the returned optimum is \(+\infty\), the original MILP is infeasible.
  Otherwise, discard the slack coordinate and solve the LP defining \(g(A\vect x)\) for the returned \(\vect x\) to recover an optimal continuous vector \(\vect y\).

  Finally, if the original LP relaxation is unbounded below, apply the preceding procedure with both objective vectors set to \(\vect 0\), starting with an LP solution for this zero objective.
  This decides mixed-integer feasibility within the same running time: if no feasible point exists, report infeasibility; otherwise, the original MILP is unbounded below.
\end{proof}

\subsection{Combinatorial n-fold}
The block-sum constraints restrict the possible solutions, which implies
\(\norm{A\vect{x}}_\infty\leq M\) where \(M=\Delta\sum_{i=1}^n b^{(i)}\) for all \(\vect x\in\bZ_{\geq 0}^t\) that adheres to the prescribed block sums.

\restate{theorem}{thm:combinatorial-nfold}

\begin{proof}
  To express the combinatorial \(n\)-fold \labelcref{eq:combinatorial-nfold} in the form required by \cref{thm:main}, define, for \(\vect{\rho}\in\ZZ^d\),
  \[
    g(\vect{\rho})=
    \begin{cases}
      f(\vect \rho), & \vect{\rho}\leq \vect{u}, \\
      +\infty,       & \text{otherwise}.
    \end{cases}
  \]
  Then \cref{eq:combinatorial-nfold} is precisely
  \cref{eq:main-problem} with this choice of \(g\). The block totals
  are already prescribed, so it remains to implement the oracle in
  \cref{eq:oracle}.

  For a given partition \(K\sqcup J=[t]\) with \(\abs{K}\leq 6d-3\) and
  \(\vect z\in\ZZ_{\geq 0}^J\), first check whether the block-sum
  constraints admit a nonnegative integer completion, reporting
  infeasibility otherwise. If they do, retain one such completion.
  The finite-valued part of the oracle is
  \begin{equation}\label{eq:combinatorial-nfold-oracle}
    \begin{aligned}
      \text{minimize}\quad   & \vect{c}_K^{\intercal}\vect{x} + f(A_K\vect x + A_J\vect z)                         \\
      \text{subject to}\quad & A_K\vect x + A_J\vect z\leq \vect{u},                                               \\
                             & \mathbf{1}^{\intercal}\vect x^{(i)} + \mathbf{1}^{\intercal}\vect z^{(i)} = b^{(i)}
      \quad(i\in[n]),                                                                                              \\
                             & \vect{x}\in\ZZ_{\geq 0}^K.
    \end{aligned}
  \end{equation}
  Introducing \(d\) integer variables for the linking image gives the
  equivalent separable convex integer program
  \begin{equation}\label{eq:combinatorial-nfold-lifted-oracle}
    \begin{aligned}
      \text{minimize}\quad   & \vect{c}_K^{\intercal}\vect{x} + f(\vect v)                                         \\
      \text{subject to}\quad & \vect v = A_K\vect x + A_J\vect z\leq \vect{u},                                     \\
                             & \mathbf{1}^{\intercal}\vect x^{(i)} + \mathbf{1}^{\intercal}\vect z^{(i)} = b^{(i)}
      \quad(i\in[n]),                                                                                              \\
                             & \vect{x}\in\ZZ_{\geq 0}^K,\quad \vect v\in\ZZ^d.
    \end{aligned}
  \end{equation}
  We solve \labelcref{eq:combinatorial-nfold-lifted-oracle} using Theorem 7.1.2 in \cite{Dadush2012}.
  Put \(N=\abs{K}+d\leq 7d-3\).
  The block sums imply \(0\leq x_j\leq b^{(i)}\) for every free variable in block \(i\), and \(-M\leq v_h\leq M\) for every \(h\in[d]\).
  These coordinate bounds have polynomial encoding length in \(\abs{I}\).

  Let \(\tilde f\) be the sum of the linear interpolations of the \(f_h\) on \([-M,M]\); it is convex and agrees with \(f\) at integer points.
  For \(\tau\in\QQ\), let \(C_\tau\) be the real relaxation of \labelcref{eq:combinatorial-nfold-lifted-oracle}, with the coordinate bounds above and \(\vect c_K^{\intercal}\vect x+\tilde f(\vect v)\leq\tau\).
  To strongly separate a query \((\overline{\vect x},\overline{\vect v})\in\QQ^N\), first check the explicit linear constraints and return any violated constraint, orienting equalities appropriately.
  Otherwise, set
  \[
    k_h=\min\set{\floor{\overline v_h},M-1},
    \qquad
    s_h=f(\vect k+\vect e_h)-f(\vect k),
  \]
  for \(h\in[d]\), where \(\vect e_h\) is the \(h\)-th unit vector.
  Separability and discrete convexity imply
  \(\tilde f(\vect v)\geq f(\vect k)+\vect s^{\intercal}(\vect v-\vect k)\)
  throughout the box, with equality at \(\overline{\vect v}\).
  If \(\vect c_K^{\intercal}\overline{\vect x}+\tilde f(\overline{\vect v})\leq\tau\), report membership in \(C_\tau\).
  Otherwise, return the valid separating inequality
  \[
    \vect c_K^{\intercal}\vect x+\vect s^{\intercal}\vect v
    \leq \tau-f(\vect k)+\vect s^{\intercal}\vect k
  \]
  which is strictly violated at the query point; a zero normal certifies \(C_\tau=\emptyset\).
  This oracle uses \(d+1\) evaluations of \(f\) at integer points in \([-M,M]^d\) and polynomial-time rational arithmetic, with separating inequalities of polynomial encoding length in \(\abs{I}\) and the encoding length of \(\tau\).
  Also, \(C_\tau\) lies in the Euclidean ball of radius \(NM\) centered at the origin.

  Dadush's deterministic integer-feasibility algorithm~\cite[Theorem~7.1.2]{Dadush2012} therefore solves each objective-threshold query in \((N+1)^{O(N)}\operatorname{poly}(\abs{I})\) time.
  Since exact polynomial-time evaluation implies polynomial encoding length for all feasible integer objective values, binary search requires only polynomially many such queries to obtain an exact minimizer.
  Thus \labelcref{eq:combinatorial-nfold-lifted-oracle} can be solved deterministically in
  \[
    Q=(N+1)^{O(N)}\operatorname{poly}(\abs{I})
    =d^{O(d)}\operatorname{poly}(\abs{I}).
  \]
  If this program is infeasible, the oracle returns \(+\infty\) together with the retained block-sum completion; otherwise, it returns an optimal \(\vect x\) and its objective value.
  The returned vector and any finite objective value have polynomial encoding length.
  Applying \cref{thm:main} now gives
  \[
    (nd\Delta)^{O(d)}
    \bigl(Q+\operatorname{poly}(\abs{I})\bigr)
    =(nd\Delta)^{O(d)}\operatorname{poly}(\abs{I}).
  \]
  If the returned optimum is \(+\infty\), the original problem is infeasible.
  Otherwise, the algorithm returns an optimal \(\vect{x}\).
\end{proof}

\subsection{Multiple Knapsack}

\restate{theorem}{thm:multiple-knapsack}

\begin{proof}[Proof sketch]
  Adding zero-profit items of weight one allows every knapsack to be filled exactly and introduces at most one additional weight class.
  Since the total available profit is fixed, we minimize rejection cost.
  For each weight class \(i\), let \(F_i(z)\) be the cost of rejecting its \(z\) cheapest items.
  Its marginal costs are nondecreasing, so \(F_i\) is discrete convex.

  We apply Rohwedder's modular construction~\cite{Rohwedder2025}.
  Knapsacks of capacity at least \(H=dw_{\max}^3\) are big; the others are small.
  We guess a pivot class \(a\) and reserve \(\abs{B}w_{\max}^2\) pivot items in total, where \(B\) is the set of big knapsacks.
  Small knapsacks are represented by full configurations.
  For big knapsacks, removing pivot items and bundles of \(w_a\) equal-weight nonpivot items leaves bounded residual configurations preserving the load modulo \(w_a\).
  A suitable pivot exists by the pigeonhole principle, and its reserve guarantees that the bundles can be distributed before inserting the pivot items.

  Grouping small knapsacks by capacity and big knapsacks by residue modulo \(w_a\) gives a combinatorial \(n\)-fold structure with \(O(H+w_a)\) configuration blocks and one bundle block.
  Applying \cref{thm:combinatorial-nfold} to each pivot gives a running time of \(w_{\max}^{O(d)}\operatorname{poly}(|I|)\).
  Appendix~\ref{app:multiple-knapsack} gives the full formulation, correctness proof, and reconstruction.
\end{proof}

\section*{Generative AI Disclosure}
ChatGPT Astra suggested algorithms for MILPs and the presented combinatorial n-fold extension using the lexicographical minimal solutions.
The authors combined them to the model presented in this paper.
ChatGPT Astra was also used to polish the text in the paper.
The authors take full responsibility for the content of the document.

\printbibliography

@inproceedings{DBLP:conf/ipco/DadushLRV23,
  author       = {Daniel Dadush and
                  Arthur L{\'{e}}onard and
                  Lars Rohwedder and
                  Jos{\'{e}} Verschae},
  title        = {Optimizing Low Dimensional Functions over the Integers},
  booktitle    = {Integer Programming and Combinatorial Optimization - 24th International
                  Conference, {IPCO} 2023, Madison, WI, USA, June 21-23, 2023, Proceedings},
  series       = {Lecture Notes in Computer Science},
  volume       = {13904},
  pages        = {115--126},
  publisher    = {Springer},
  year         = {2023},
  doi          = {10.1007/978-3-031-32726-1_9},
  bibsource    = {dblp computer science bibliography, https://dblp.org}
}

@article{Lenstra1983,
  author  = {Lenstra, Jr., Hendrik W.},
  title   = {Integer Programming with a Fixed Number of Variables},
  journal = {Math. Oper. Res.},
  volume  = {8},
  number  = {4},
  pages   = {538--548},
  year    = {1983},
  doi     = {10.1287/moor.8.4.538}
}

@article{Kannan1987,
  author  = {Ravi Kannan},
  title   = {Minkowski's Convex Body Theorem and Integer Programming},
  journal = {Math. Oper. Res.},
  volume  = {12},
  number  = {3},
  pages   = {415--440},
  year    = {1987},
  doi     = {10.1287/moor.12.3.415}
}

@phdthesis{Dadush2012,
  author             = {Daniel Nicolas Dadush},
  title              = {Integer Programming, Lattice Algorithms, and Deterministic Volume Estimation},
  school             = {Georgia Institute of Technology},
  year               = {2012},
  timestamp          = {Tue, 29 Apr 2025 01:00:00 +0200},
  biburl             = {https://dblp.org/rec/phd/basesearch/Dadush12.bib},
  bibsource          = {dblp computer science bibliography, https://dblp.org},
  url                = {http://hdl.handle.net/1853/44807},
  _bib2doi_selected  = {dblp:/rec/phd/basesearch/Dadush12.bib},
  _bib2doi_confirmed = {true},
  _bib2doi_finished  = {true}
}

@inproceedings{Rohwedder2025,
  author    = {Lars Rohwedder},
  title     = {{ETH}-Tight {FPT} Algorithm for Makespan Minimization on Uniform Machines},
  booktitle = {52nd International Colloquium on Automata, Languages, and Programming (ICALP 2025)},
  series    = {Leibniz International Proceedings in Informatics (LIPIcs)},
  volume    = {334},
  pages     = {126:1--126:13},
  publisher = {Schloss Dagstuhl -- Leibniz-Zentrum f{\"u}r Informatik},
  year      = {2025},
  doi       = {10.4230/LIPIcs.ICALP.2025.126},
}

@article{EisenbrandWeismantel2020,
  author  = {Friedrich Eisenbrand and Robert Weismantel},
  title   = {Proximity Results and Faster Algorithms for Integer Programming
             Using the {Steinitz} Lemma},
  journal = {ACM Transactions on Algorithms},
  volume  = {16},
  number  = {1},
  pages   = {5:1--5:14},
  year    = {2020},
  doi     = {10.1145/3340322}
}

@article{JansenRohwedder2023,
  author  = {Klaus Jansen and Lars Rohwedder},
  title   = {On Integer Programming, Discrepancy, and Convolution},
  journal = {Mathematics of Operations Research},
  volume  = {48},
  number  = {3},
  pages   = {1481--1495},
  year    = {2023},
  doi     = {10.1287/moor.2022.1308}
}

@inproceedings{CslovjecsekEtAl2021,
  author    = {Jana Cslovjecsek and Friedrich Eisenbrand and
               Christoph Hunkenschr{\"o}der and Lars Rohwedder and
               Robert Weismantel},
  title     = {Block-Structured Integer and Linear Programming in Strongly
               Polynomial and Near Linear Time},
  booktitle = {Proceedings of the 2021 ACM-SIAM Symposium on Discrete
               Algorithms ({SODA})},
  pages     = {1666--1681},
  publisher = {SIAM},
  year      = {2021},
  doi       = {10.1137/1.9781611976465.101}
}

@inproceedings{BrandEtAl2024,
  author    = {Cornelius Brand and Martin Kouteck{\'y} and
               Alexandra Lassota and Sebastian Ordyniak},
  title     = {Separable Convex Mixed-Integer Optimization:
               Improved Algorithms and Lower Bounds},
  booktitle = {32nd Annual European Symposium on Algorithms ({ESA} 2024)},
  series    = {Leibniz International Proceedings in Informatics (LIPIcs)},
  volume    = {308},
  pages     = {32:1--32:18},
  publisher = {Schloss Dagstuhl -- Leibniz-Zentrum f{\"u}r Informatik},
  year      = {2024},
  doi       = {10.4230/LIPIcs.ESA.2024.32},
}

@article{KnopKouteckyMnich2020,
  author  = {Du{\v{s}}an Knop and Martin Kouteck{\'y} and Matthias Mnich},
  title   = {Combinatorial {$n$}-fold Integer Programming and Applications},
  journal = {Mathematical Programming},
  volume  = {184},
  number  = {1--2},
  pages   = {1--34},
  year    = {2020},
  doi     = {10.1007/s10107-019-01402-2}
}

@inproceedings{JansenEtAl2025,
  author    = {Klaus Jansen and Kai Kahler and Lis Pirotton and Malte Tutas},
  title     = {New Algorithm for Combinatorial {$n$}-Folds and Applications},
  booktitle = {20th International Symposium on Parameterized and Exact
               Computation ({IPEC} 2025)},
  series    = {Leibniz International Proceedings in Informatics (LIPIcs)},
  volume    = {358},
  pages     = {15:1--15:15},
  publisher = {Schloss Dagstuhl -- Leibniz-Zentrum f{\"u}r Informatik},
  year      = {2025},
  doi       = {10.4230/LIPIcs.IPEC.2025.15},
}

@inproceedings{GuptaEtAl2025,
  author    = {Sushmita Gupta and Pallavi Jain and Sanjay Seetharaman and
               Meirav Zehavi},
  title     = {A Simple Algorithm for Combinatorial {$n$}-Fold {ILPs}
               Using the {Steinitz} Lemma},
  booktitle = {20th International Symposium on Parameterized and Exact
               Computation ({IPEC} 2025)},
  series    = {Leibniz International Proceedings in Informatics (LIPIcs)},
  volume    = {358},
  pages     = {14:1--14:17},
  publisher = {Schloss Dagstuhl -- Leibniz-Zentrum f{\"u}r Informatik},
  year      = {2025},
  doi       = {10.4230/LIPIcs.IPEC.2025.14},
}

@article{KouteckyZink2025,
  author  = {Martin Kouteck{\'y} and Johannes Zink},
  title   = {Complexity of Scheduling Few Types of Jobs on Related
             and Unrelated Machines},
  journal = {Journal of Scheduling},
  volume  = {28},
  number  = {1},
  pages   = {139--156},
  year    = {2025},
  doi     = {10.1007/s10951-024-00827-8}
}

@inproceedings{ChenJansenZhang2014,
  author    = {Lin Chen and Klaus Jansen and Guochuan Zhang},
  title     = {On the Optimality of Approximation Schemes for the
               Classical Scheduling Problem},
  booktitle = {Proceedings of the Twenty-Fifth Annual ACM-SIAM Symposium
               on Discrete Algorithms ({SODA} 2014)},
  pages     = {657--668},
  publisher = {SIAM},
  year      = {2014},
  doi       = {10.1137/1.9781611973402.50},
}

@article{StearnsHunt1990,
  author  = {Richard Edwin Stearns and Hunt, III, Harry B.},
  title   = {Power Indices and Easier Hard Problems},
  journal = {Mathematical Systems Theory},
  volume  = {23},
  number  = {4},
  pages   = {209--225},
  year    = {1990},
  doi     = {10.1007/BF02090776},
}

@article{GurskiRehsRethmann2019,
  author  = {Frank Gurski and Carolin Rehs and Jochen Rethmann},
  title   = {Knapsack Problems: A Parameterized Point of View},
  journal = {Theoretical Computer Science},
  volume  = {775},
  pages   = {93--108},
  year    = {2019},
  doi     = {10.1016/j.tcs.2018.12.019}
}

@article{ChekuriKhanna2005,
  author  = {Chandra Chekuri and Sanjeev Khanna},
  title   = {A Polynomial Time Approximation Scheme for the Multiple
             Knapsack Problem},
  journal = {SIAM Journal on Computing},
  volume  = {35},
  number  = {3},
  pages   = {713--728},
  year    = {2005},
  doi     = {10.1137/S0097539700382820}
}

@article{Jansen2009,
  author  = {Klaus Jansen},
  title   = {Parameterized Approximation Scheme for the Multiple
             Knapsack Problem},
  journal = {SIAM Journal on Computing},
  volume  = {39},
  number  = {4},
  pages   = {1392--1412},
  year    = {2009},
  doi     = {10.1137/080731207}
}

@inproceedings{Jansen2012,
  author    = {Klaus Jansen},
  title     = {A Fast Approximation Scheme for the Multiple Knapsack Problem},
  booktitle = {{SOFSEM} 2012: Theory and Practice of Computer Science},
  series    = {Lecture Notes in Computer Science},
  volume    = {7147},
  pages     = {313--324},
  publisher = {Springer},
  year      = {2012},
  doi       = {10.1007/978-3-642-27660-6_26}
}

@article{DBLP:journals/siamjo/HunkenschroderP23,
  author       = {Christoph Hunkenschr{\"{o}}der and
                  Sebastian Pokutta and
                  Robert Weismantel},
  title        = {Minimizing a Low-Dimensional Convex Function Over a High-Dimensional
                  Cube},
  journal      = {{SIAM} J. Optim.},
  volume       = {33},
  number       = {2},
  pages        = {538--552},
  year         = {2023},
  doi          = {10.1137/22M1489988},
  bibsource    = {dblp computer science bibliography, https://dblp.org}
}

@phdthesis{DBLP:phd/dnb/Rohwedder19,
  author       = {Lars Rohwedder},
  title        = {Algorithms for Integer Programming and Allocation},
  school       = {University of Kiel, Germany},
  year         = {2019},
  url          = {https://macau.uni-kiel.de/receive/dissertation_diss_00026125},
  urn          = {urn:nbn:de:gbv:8-diss-261258},
  bibsource    = {dblp computer science bibliography, https://dblp.org}
}

@article{DBLP:journals/mp/Meyer74,
  author       = {Robert R. Meyer},
  title        = {On the existence of optimal solutions to integer and mixed-integer
                  programming problems},
  journal      = {Math. Program.},
  volume       = {7},
  number       = {1},
  pages        = {223--235},
  year         = {1974},
  doi          = {10.1007/BF01585518},
  bibsource    = {dblp computer science bibliography, https://dblp.org}
}

@article{DBLP:journals/mp/CookGST86,
  author       = {William J. Cook and
                  A. M. H. Gerards and
                  Alexander Schrijver and
                  {\'{E}}va Tardos},
  title        = {Sensitivity theorems in integer linear programming},
  journal      = {Math. Program.},
  volume       = {34},
  number       = {3},
  pages        = {251--264},
  year         = {1986},
  doi          = {10.1007/BF01582230},
  bibsource    = {dblp computer science bibliography, https://dblp.org}
}

@inproceedings{DBLP:conf/ipco/RohwedderW25,
  author       = {Lars Rohwedder and
                  Karol W{\k{e}}grzycki},
  title        = {Space-Efficient Algorithm for Integer Programming with Few Constraints},
  booktitle    = {Integer Programming and Combinatorial Optimization - 26th International
                  Conference, {IPCO} 2025, Baltimore, MD, USA, June 11-13, 2025, Proceedings},
  series       = {Lecture Notes in Computer Science},
  volume       = {15620},
  pages        = {442--451},
  publisher    = {Springer},
  year         = {2025},
  doi          = {10.1007/978-3-031-93112-3_32},
  bibsource    = {dblp computer science bibliography, https://dblp.org}
}

@article{DBLP:journals/mp/KnopKLMO23,
  author       = {Dusan Knop and
                  Martin Kouteck{\'{y}} and
                  Asaf Levin and
                  Matthias Mnich and
                  Shmuel Onn},
  title        = {High-multiplicity N-fold {IP} via configuration {LP}},
  journal      = {Math. Program.},
  volume       = {200},
  number       = {1},
  pages        = {199--227},
  year         = {2023},
  doi          = {10.1007/S10107-022-01882-9},
  bibsource    = {dblp computer science bibliography, https://dblp.org}
}

\clearpage
\appendix
\section{Full Proof for Multiple Knapsack}\label{app:multiple-knapsack}

Since the total profit \(\sum_i p_i n_i\) of all available items is
fixed, maximizing the profit of selected items is equivalent to
minimizing the total profit of the items not selected. We regard the
profit of an unselected item as its rejection cost. Items of the same
weight can then be grouped: if a prescribed number of them is to be
rejected, it is optimal to reject the cheapest ones. This gives a
separable convex rejection cost. We combine this observation with the
modular pivot construction of Rohwedder~\cite{Rohwedder2025} and
\cref{thm:combinatorial-nfold}.

Put \(T=\sum_{j=1}^m C_j\).
We assume \(w_{\max}\geq2\), since otherwise selecting the most profitable items is optimal.
Add \(T\) dummy items of weight one and profit zero.
Every packing can now be filled to capacity without changing its profit, and discarding dummy items recovers a packing of the original instance.
Group the augmented items by weight, merging the dummies with any real items of weight one.
Adding the dummies introduces at most one new weight class, which is absorbed in the \(w_{\max}^{O(d)}\) running-time bound.
Thus, without loss of generality, we continue to write \(d\) for the number of weight classes after augmentation.
Index these classes by their distinct weights \(w_1,\ldots,w_d\), with multiplicities \(N_1,\ldots,N_d\).
In particular, \(d\leq w_{\max}\).
For \(i\in[d]\) and \(z\in\{0,\ldots,N_i\}\), let \(F_i(z)\) be the total rejection cost of the \(z\) cheapest items of weight \(w_i\).
The differences \(F_i(z+1)-F_i(z)\) are nondecreasing, so \(F_i\) is discrete convex.
Sorting the profit classes within each weight class gives a piecewise-linear interpolation with one segment per distinct profit.
Its breakpoints are prefix sums of binary multiplicities.
Thus the functions have polynomial encoding length and can be evaluated exactly in polynomial time.
Extend them beyond their domains using their first and last segment slopes; these extensions remain convex.

To formulate the packing constraints, use Rohwedder's modular construction~\cite{Rohwedder2025} and set
\[
  H=dw_{\max}^3,\qquad B=\{j:C_j\geq H\},\qquad R=|B|w_{\max}^2.
\]
Call the knapsacks in \(B\) big and all others small.
For each candidate pivot class \(a \in [d]\) satisfying \(N_a\geq R\), solve the formulation below.
We take the best solution over all candidates.
The formulation will reserve \(R\) items of this class for the big knapsacks.
If \(B=\emptyset\), then \(R=0\).

For a fixed pivot \(a\), group small knapsacks by their exact capacity and big knapsacks by their capacity modulo \(w_a\).
Let \(\mathcal T\) be the set of nonempty classes and \(m_\tau\) the number of knapsacks in class \(\tau \in \mathcal T\).
There are at most \(H+w_a\) classes.
A small class of capacity \(t<H\) has configurations
\[
  \mathcal C_\tau=
  \{\vect{c}\in\ZZ_{\geq0}^d:\vect{w}^{\intercal}\vect{c}=t\}.
\]
For a big class of residue \(\rho\), use the residual configurations
\[
  \mathcal C_\tau=
  \left\{\vect{c}\in\ZZ_{\geq0}^d:
  \begin{array}{l}
    c_a=0,\quad c_i<w_a\text{ for }i\ne a, \\
    \vect{w}^{\intercal}\vect{c}\equiv\rho\pmod{w_a}
  \end{array}\right\}.
\]
A bundle of a nonpivot class contains \(w_a\) items, whereas a pivot bundle contains one item. Write \(\gamma_i=w_a\) for \(i\ne a\) and \(\gamma_a=1\).

Let \(x_{\tau,\vect{c}}\) count knapsacks using configuration \(\vect{c}\), and let \(b_i\) count bundles of class \(i\), excluding the reserved pivot items.
The selected count outside the reserve is
\[
  Y_i=\sum_{\tau\in\mathcal T}\sum_{\vect{c}\in\mathcal C_\tau}
  c_i x_{\tau,\vect{c}}+\gamma_i b_i.
\]
Consequently, the number of rejected items for a given class \(i\) is
\[
  N_i- ( Y_i + R\,1_{i=a} ) = \begin{cases}
    N_i - Y_i - R, & i=a,    \\
    N_i - Y_i,     & i\ne a.
  \end{cases}
\]
We obtain the minimization problem
\begin{equation}\label{eq:multiple-knapsack}
  \begin{aligned}
    \min\quad & \sum_{i=1}^d F_i(N_i- Y_i - R\,1_{i=a})                                                       \\
    \text{subject to}\quad
              & \sum_{\vect{c}\in\mathcal C_\tau}x_{\tau,\vect{c}}=m_\tau
              &                                                           & (\tau\in\mathcal T),              \\
              & Y_i\leq N_i-R\,1_{i=a}                                    &                      & (i\in[d]), \\
              & \sum_{i=1}^d w_iY_i\leq T-w_a R,                                                              \\
              & \sum_{i=1}^d b_i+h=T,                                                                         \\
              & x_{\tau,\vect{c}},b_i,h\in\ZZ_{\geq0}.
  \end{aligned}
\end{equation}
The last equality creates one additional block containing the bundle variables and a slack variable \(h\).
It does not restrict the other variables: every bundle has weight at least one, so the total-weight inequality already implies \(\sum_i b_i\leq T\).
The reserve is accounted for both in the selected weight and in the rejection costs.

The correctness of this formulation follows essentially the same argument as Rohwedder~\cite[Sections~2 and~4]{Rohwedder2025}.

\begin{lemma}\label{lem:multiple-knapsack-formulation}
  For every \(\kappa\in\QQ_{\geq0}\), the following are equivalent:
  \begin{enumerate}
    \item The original instance admits a packing with rejection cost
          at most \(\kappa\).
    \item For some pivot class \(a\in[d]\) satisfying \(N_a\geq R\),
          \cref{eq:multiple-knapsack} admits a feasible solution with
          objective value at most \(\kappa\).
  \end{enumerate}
  Moreover, any feasible solution of
  \cref{eq:multiple-knapsack} can be converted into a packing
  with the same rejection cost in time polynomial in its
  encoding length and \(|I|\).
\end{lemma}

\begin{proof}
  Take an optimal packing and fill its unused capacity with dummy items.
  Every big knapsack contains at least \(H/w_{\max}=dw_{\max}^2\) items, so by the pigeonhole principle, some weight class contributes at least \(R=|B|w_{\max}^2\) items to the big knapsacks.
  Choose this class as the pivot; if there are no big knapsacks, choose any class.
  Keep the full configurations of the small knapsacks.
  From every big knapsack remove all pivot items and groups of \(w_a\) nonpivot items of the same weight until fewer than \(w_a\) of each nonpivot class remain.
  These removals preserve the load modulo \(w_a\).
  Reserve \(R\) of the removed pivot items and record all other removed items as bundles.
  This gives a feasible solution of \cref{eq:multiple-knapsack}.
  Within each weight class, the rejected items may be chosen to be the cheapest ones without changing feasibility.
  Hence the objective equals the rejection cost of an optimal packing.

  Conversely, assign the configurations of a feasible solution to knapsacks of their classes. Small knapsacks are full.
  Every big residual configuration fits, since \(\vect{w}^{\intercal}\vect{c}\leq(d-1)(w_a-1)w_{\max}<H\).
  Keep the \(R+b_a\) pivot items aside and insert the nonpivot bundles first.
  At every stage, the total free space in the big knapsacks is at least the weight of all selected but unplaced items.
  In particular, it is at least \(w_a R\geq |B|w_{\max}^2\). If \(B\ne\emptyset\), some big knapsack therefore has free space at least \(w_{\max}^2\).
  Every nonpivot bundle has weight \(w_a w_i\leq w_{\max}^2\) and can therefore be placed. Afterwards, all remaining free spaces are multiples of \(w_a\) and their sum is at least \(w_a(R+b_a)\).
  Thus all pivot items can also be placed.
  If \(B=\emptyset\), the small configurations already have total weight \(T\), so the weight inequality forces all bundle variables to zero.

  Choose the rejected items in each class to be its cheapest items, assign the remaining copies to the constructed positions of that weight, and discard dummy items.
  This produces a packing with rejection cost exactly the objective of \cref{eq:multiple-knapsack}.
  Taking the minimum over pivot choices therefore yields an optimal solution of the original instance.
  Bundle placement and the assignment of original profit classes can both be performed in batches, so reconstruction is polynomial in the input and output encoding lengths.
\end{proof}

\begin{proof}[of \cref{thm:multiple-knapsack}]
  By \cref{lem:multiple-knapsack-formulation}, it remains to solve \cref{eq:multiple-knapsack}, which we do by applying \cref{thm:combinatorial-nfold}.
  Use the \(d+1\) inequalities involving \(Y_i\) as linking rows, and define
  \[
    f_i(t)=F_i(N_i-R\,1_{i=a}-t)\quad(i\in[d]),
    \qquad f_{d+1}(t)=0.
  \]
  Each \(f_i\) is discrete convex.
  The polynomial-time exact evaluation assumption of \cref{thm:combinatorial-nfold} holds by the piecewise-linear descriptions above.
  The formulation has \(d+1\) linking rows and at most \(H+w_a+1\) blocks, with coefficient bound \(\Delta\leq H\leq w_{\max}^4\).
  There are at most \((H+w_a)(H+1)^d=w_{\max}^{O(d)}\) configurations.
  Their encoding, together with the rejection-cost functions, has length \(w_{\max}^{O(d)}\cdot \operatorname{poly}(|I|)\); the number of profit classes affects only the polynomial factor.

  Since \(d\leq w_{\max}\) and \(w_a\leq w_{\max}\), there are at most \(2w_{\max}\) linking rows and \(3w_{\max}^4\) blocks.
  Applying \cref{thm:combinatorial-nfold} to each of the \(d\) pivot choices therefore takes
  \[
    w_{\max}^{O(d)}\cdot \operatorname{poly}(|I|)
  \]
  time, including configuration enumeration and reconstruction.
\end{proof}
\end{document}